\documentclass[nofootinbib,preprint,aps,draft,pra,superscriptaddress,citeautoscript,floatfix]{revtex4-1}
\setcitestyle{super}
\usepackage{graphics}
\usepackage{graphicx}
\usepackage{color}
\usepackage{amsmath}
\usepackage{amssymb}
\usepackage{float}
\newcommand\be{\begin{equation}}
\newcommand\ee{\end{equation}}
\newcommand\bea{\begin{eqnarray}}
\newcommand\eea{\end{eqnarray}}
\begin{document}

\title{Enhanced electro-actuation in dielectric elastomers: the non-linear effect of free ions}
\author{Bin Zheng}
\affiliation{Raymond and Beverly Sackler School of Physics and Astronomy,
Tel Aviv University, Ramat Aviv 69978, Tel Aviv, Israel}
\author{Xingkun Man}
\email{manxk@buaa.edu.cn}
\affiliation{Center of Soft Matter Physics and its Applications, and School of
Physics and Nuclear Energy Engineering, Beihang University, Beijing 100191, China}
\author{David Andelman}
\email{andelman@post.tau.ac.il}
\affiliation{Raymond and Beverly Sackler School of Physics and Astronomy,
Tel Aviv University, Ramat Aviv 69978, Tel Aviv, Israel}
\author{Masao Doi}
\email{masao.doi@buaa.edu.cn}
\affiliation{Center of Soft Matter Physics and its Applications, and School of
Physics and Nuclear Energy Engineering, Beihang University, Beijing 100191, China}

\begin{abstract}
Plasticized poly(vinyl chloride) (PVC) is a jelly-like soft dielectric material that attracted substantial interest recently as a new type of electro-active polymers.
Under electric fields of several hundred Volt/mm, PVC gels undergo large deformations. These gels can be used as artificial muscles and other soft robotic devices, with striking deformation behavior that is quite different from conventional dielectric elastomers.
Here, we present a simple model for the electro-activity of PVC gels, and show a non-linear effect of free ions on its dielectric behaviors. It is found that their particular deformation behavior is due to an electro-wetting effect and to a change in their interfacial tension. In addition, we derive analytical expressions for the surface tension as well as for the apparent dielectric constant of the gel. The theory indicates that the size of the mobile free ions has a crucial role in determining the electro-induced deformation, opening up the way to novel and innovative designs of electro-active gel actuators.
\end{abstract}

\maketitle

Polyvinyl chloride (PVC) mixed with plasticizers form a jelly-like soft gel dielectrics, and have attracted growing interest in recent years as soft actuators\cite{Hirai2001,Asaka2014}.
Compared with conventional dielectric elastomers\cite{Hines2017, suo2016, suo2018, Erbas2014,pei2010,Fuchigami2014, Carpi2011, Romasanta2015},
plasticized PVC gels offer several advantages in terms of their fast response, low working voltage, larger sustained
stress, and overall stability and durability\cite{Hirai2001,Asaka2014,Yi2019}.
Such properties make these materials attractive for novel applications in soft robotics,
artificial muscles, medical assistance devices\cite{li2015,Ko2018}
and in electro-optical devices \cite{Bae2017,Hiromu2013}.

\begin{figure}[h]
\centering{\includegraphics[width=0.7\textwidth,draft=false]{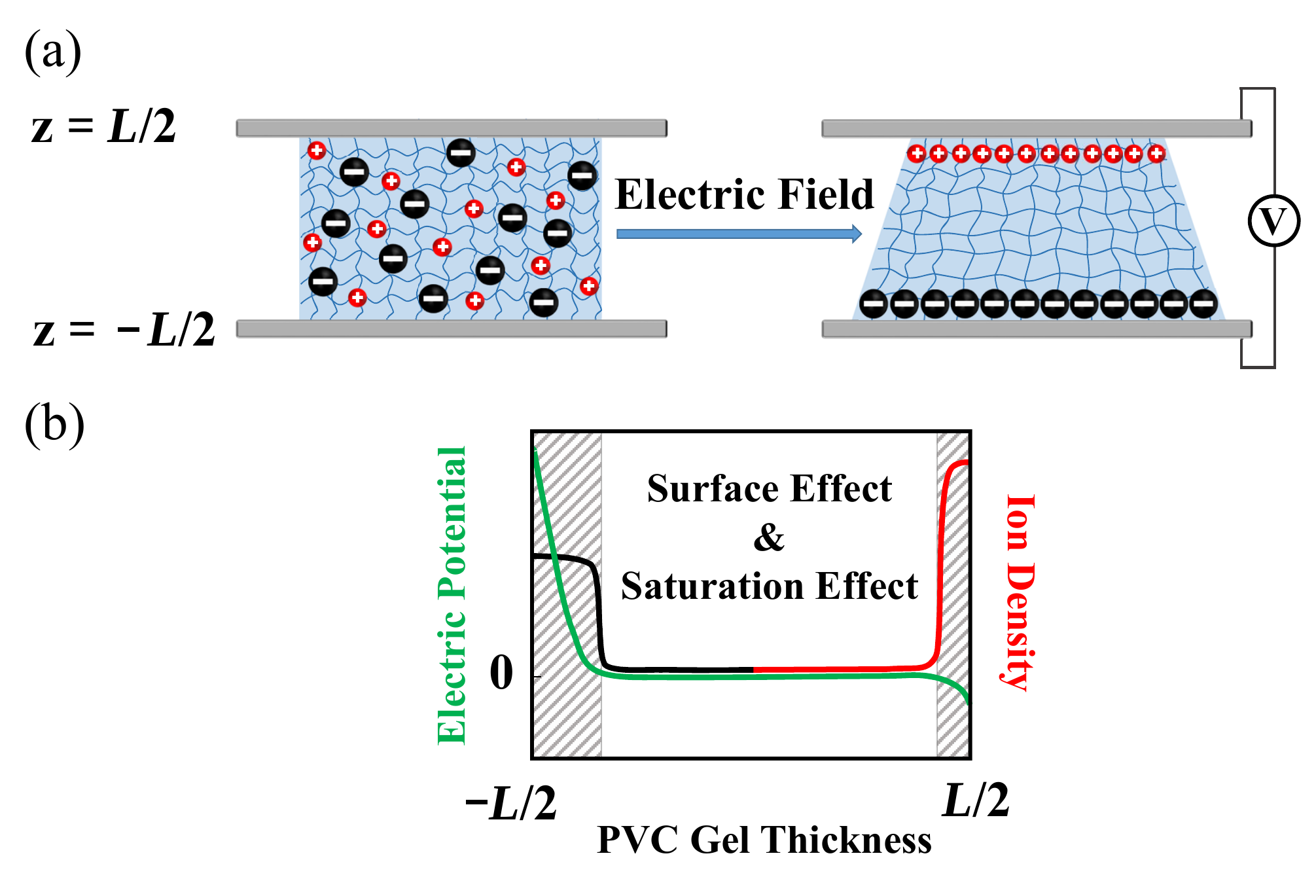}}
\caption{
\textsf{Schematic drawing of the asymmetric deformation of the PVC gel in an electric field and the electrostatic profiles as a function of the interplate distance. (a) The PVC gel is placed between two flat and parallel electrodes located at $z=\pm L/2$. The charge density on the anode and cathode is $q_\pm=\pm Q/S_\pm$, respectively, where $S_\pm$ is the contact area, and the total charge on each of the two electrodes is $\pm Q$ (in units of e). Before applying the electric field, the cations and anions inside the PVC gel are homogeneously distributed.  (b) Schematic plots of the ion distribution and electrostatic potential inside the PVC gel. The electric field in the bulk of the gel film is nearly zero, and the large electric field exists only near the electrodes, where a very dense layer of free ions is created.
}}
\label{fig1}
\end{figure}

PVC gels are found to have quite different properties than conventional dielectric elastomers~\cite{Hirai1999,Ali2011,Ali2012,xia2010,xia2010a}.
First, their apparent dielectric constant is $10^3$-$10^4$ times larger than in
conventional materials. Second, upon the application of an electric
field, the PVC gel shows amoeba-like spreading on the anode (and not on the cathode), as is shown schematically in Fig.~\ref{fig1}. This behavior is quite unique and different from the gel contraction observed under similar conditions in most dielectric elastomers\cite{suo2018, Romasanta2015}. Note that when the polarity of the applied voltage is reversed, the PVC gel deformation will also change its direction. This finding was
explained by proposing that the electrostatic adhesive force generated by
charges accumulates at the anode~\cite{Asaka2018}. The injected charges from the cathode are important as there will be no deformation if the cathode is blocked by coating it with an insulating film.\cite{Ali2012}
Furthermore, a thin plasticizer-rich layer was found experimentally to coat the anode
(but not the cathode)~\cite{xia2010}. It has been conjectured~\cite{xia2010a,Ali2012,Asaka2018} that such characteristic behavior is due to the existence of small amounts of free ions in the PVC gel, but to the best of our knowledge, no available model exists that can predict such a material deformation for experimental conditions, and eventually to facilitate design of actuators in device applications.

In this Letter, we propose an electro-wetting mechanism to explain the PVC gel electro-activity. The deformation of dielectric elastomers (DEAs) conventionally originates from the Coulomb attraction between the electrodes, {\it i.e.} the Maxwell stress. However, the latter is a bulk effect, while our interfacial effect has a different origin. The voltage drop we consider is of the order of hundreds of volts~\cite{Ali2011}, making it much larger than in regular electrolytes (few volts)\cite{david1997,david2000}. Only at the electrode proximity, a very concentrated layer of free ions is established, where the ionic saturation effect is considered. We shall show that the ionic size asymmetry creates a large voltage drop at the anode side, yielding a negative contribution to the interfacial tension between the gel and anode, and hence leading to an instability at the anode with amoeba-like deformation of the gel.

We consider a set-up where a PVC gel is confined between two flat electrodes located at $z=\pm L/2$. The gel is an insulator with a dielectric constant $\varepsilon_r$.
We further assume that the free charges that are embedded in the gel have an asymmetric size, $a_{-}>a_{+}$, where $a_{\pm}$ is defined as the cubic root of the ion molecular volume and is related to the close-packing density, $1/a_{\pm}^3$. The two electrodes have fixed and opposite total charge $\pm Q$ (in units of $e$), yielding a charge density $q_\pm=\pm Q/S_\pm$ at the anode and cathode, respectively, where $S_\pm$ is the contact area of the gel with the corresponding electrode. As $S_\pm$ can vary as the deformation occurs, so will $q_\pm$.

Next we derive separately the electrostatic free energy and elastic energy associated with the confined gel. As the voltages applied in the experiment are high (on the order of $10^3{\rm V}$), we employ the saturation condition of the free-ion density in order to avoid any nonphysical accumulation of ions near the electrodes (see Fig.~\ref{fig1}b). The electrostatic free energy was proposed in Refs. \cite{david1997,david2000} and is written as,
\be
\label{e1}
\begin{aligned}
\frac{F_{{\rm elec}}}{k_{\rm B}T}=&\int (f-f_{\rm bulk})\,{\rm d}^3r   \\
=&\int{\rm d^3}r \left\{-\frac{1}{8\pi l_{\rm B}}(\nabla\Psi )^2+(c_{+}-c_{-})\Psi
\right. \\
&\left. -\tilde\mu_{+}c_{+}-\tilde\mu_{-}c_{-}+c_{+}{\rm ln}(c_{+}a_{+}^3)+c_{-}{\rm ln}(c_{-}a_{-}^3)  \right. \\
&+a_{+}^{-3}(1-c_{+}a_{+}^3){\rm ln}(1-c_{+}a_{+}^3)  \\
&+a_{-}^{-3}(1-c_{-}a_{-}^3){\rm ln}(1-c_{-}a_{-}^3) \bigg\}
\end{aligned}
\ee
where the first two terms account for the electrostatic energy, the following two terms account for ionic chemical potential, and the last four terms account for the entropy of mixing, where the hard-core nature of the ions is included through the last two terms. $l_{\rm B}=e^2/(4\pi\varepsilon_0\varepsilon_rk_BT)$ is the Bjerrum length, $\Psi={e\psi}/{k_{\rm B}T}$ is the dimensionless electrostatic potential, $c_\pm(r)$ is the monovalent cation and anion number density, and $\tilde\mu_\pm={\mu_\pm}/{k_{\rm B}T}$ is the dimensionless chemical potential. In our model, the ions interact with the polymer matrix but the interaction does not appear explicitly in the present theory. The free energy form of Eq.~(\ref{e1}) indicates the interaction between ions and polymer matrix is implicit; namely, the interaction is included in the dielectric constant $\varepsilon$ and in the incompressible condition of the system (ions and polymer chains). More details on Eq.~(\ref{e1}) and the reason behind this simplified form of its entropy terms are given in the supplemental material.

The variation of the free energy $F_{\rm elec}$ with
respect to $\Psi$ and $c_\pm$ yields a modified Poisson-Boltzmann equation\cite{david1997,david2000} that includes the steric effects,

\be
\label{e2}
\begin{aligned}
\Psi''(z)=-\frac{\kappa_{\rm D}^2}{2}\left(\frac{{\rm e}^{-\Psi(z)}}{1-\phi_{+}^b+\phi_{+}^b{\rm e}^{-\Psi(z)}}-\frac{{\rm e}^{\Psi(z)}}{1-\phi_{-}^b+\phi_{-}^b{\rm e}^{\Psi(z)}}\right)   ,
\end{aligned}
\ee
where the ionic volume fraction is $\phi_{\pm}^b\equiv c_b a_\pm^3$ and the Debye length is
$\lambda_{\rm D}=1/\kappa_{\rm D}=1/\sqrt{8\pi l_{\rm B} c_b}$, where $c_b$ is the bulk free ion concentration. The boundary condition at $z=\pm L/2$ is $\Psi'(\pm L/2)=\pm 4\pi l_{\rm B} q_\mp$. In accordance with the experimental conditions, we model the high voltage regime by choosing large values of the surface charge density, $q_\pm$. Moreover, the thickness $L$ of the confined gel, which is of the order of millimeters, is much larger than the Debye length (on the order of nanometers), $L\gg \lambda_{\rm D}$. Hence, the electrostatic potential is vanishingly small in the bulk of the gel, $\Psi(z)\approx 0$, and the free-ion densities there are $c_{+}(z)\approx c_{-}(z)\approx c_b$.
This means that the two charged boundaries are decoupled, and one can focus on half of the system with a new boundary condition introduced at the symmetric mid-plane ($z = 0$) with $\Psi(0) = 0$, $\Psi'(0) = 0$, and $c_{+}(0) = c_{-}(0) \approx c_b$.

The excess free-energy for $-L/2\le z \le 0$ (per unit area) can be written as,
\be
\label{e3}
\begin{aligned}
g(q_+)=&\int_{-L/2}^{0}{\rm d}z\left[f(c_{+}(z),c_{-}(z),\Psi(z))-f(c_b,c_b,0)\right]\\
&+ q_+ \Psi(-L/2) ,
\end{aligned}
\ee
where $f$ is the free-energy density in Eq.~(\ref{e1}), and the last term of $g$ accounts for the electrostatic energy of the electrode.

We assume that the surface tension is homogenous at the electrodes and consider the spreading of the gel at the electrode by analyzing the changes in the contact area at each of the two electrodes. The total surface free energy is obtained by multiplying the contact area at the electrodes with the calculated free-energy density. If an electrode has an area $S_+$, and charge $Q$, the surface free energy of the anode is given by $S_+[\gamma_0+g(Q/S_+)]$, where $\gamma_0$ is the surface tension when $Q=0$.  The surface tension $\gamma_+$ of the anode is then given by

\be
\label{e4}
\begin{aligned}
 \Delta\gamma_{+}(q_+)=&\frac{\partial [S_+ g(Q/S_+)]}{\partial S_+} \\
 =&g(q_+)-q_+\frac{\partial g(q_+)}{\partial q_+}.
\end{aligned}
\ee

Similarly, $g(q_-)$ and $\gamma_{-}(q_-)$ is the excess free energy and the surface tension between the PVC gel and the cathode, respectively, for the other half film $[0, L/2]$.

Finally, the mechanical deformation of the gel is given by the balance in the full free-energy between the excess free-energy and the gel elastic-energy,
\be
\label{e5}
\begin{aligned}
G=S_{+}g(Q/S_{+})+S_{-}g(-Q/S_{-})+F_{{\rm el}}(S_{+},S_{-})
\end{aligned}
\ee
where $F_{{\rm el}}(S_{+},S_{-})$ is a phenomenological gel elastic-energy given by
\be
\label{e6}
\begin{aligned}
\frac{F_{{\rm el}}}{k_{\rm B}T}=\frac{k_b}{2}(S_{+}-S_0)^2+\frac{k_b}{2}(S_{-}-S_0)^2+\frac{k_s}{2}(S_{+}-S_{-})^2
\end{aligned}
\ee
where $S_0$ is the initial (no electric field) contact area of the gel at both electrodes. The two elastic constants appearing above are $k_b={K_b}/{k_{\rm B}T}$ and $k_s={K_s}/{k_{\rm B}T}$. In this phenomenological elastic energy, we have integrated the contribution from the profile of the whole (bulk) gel deformation, in addition to the surface contribution, yielding a deformation of a thin gel film under an electric field. This elastic deformation of the gel leads to a change in the contact area. As this gel deformation is very complex, we take instead a phenomenological approach. We use the change in the contact area to characterize the gel elastic deformation by introducing two phenomenological elastic constants, $k_b$ and $k_s$ that account for the whole bulk deformation. The non-uniform distribution of ions may affect the elasticity of the matrix as the excluded volume of the ions is taken explicitly into account in our theory. However, in the present work, the ionic layer formed by the accumulation of free ions near the electrodes is very thin (about $80{\rm \AA}$), as compared with the thickness of the gel film (about $1$mm). Therefore, we assume that such a thin layer will not affect substantially the elasticity of the soft gel film. Then, the full free-energy can be expressed as a function of $S_{+}$ and $S_{-}$, and the gel deformation is determined by minimizing $G$ with respect to $S_{+}$ and $S_{-}$.

\begin{figure}[h]
{\includegraphics[width=0.6\textwidth,draft=false]{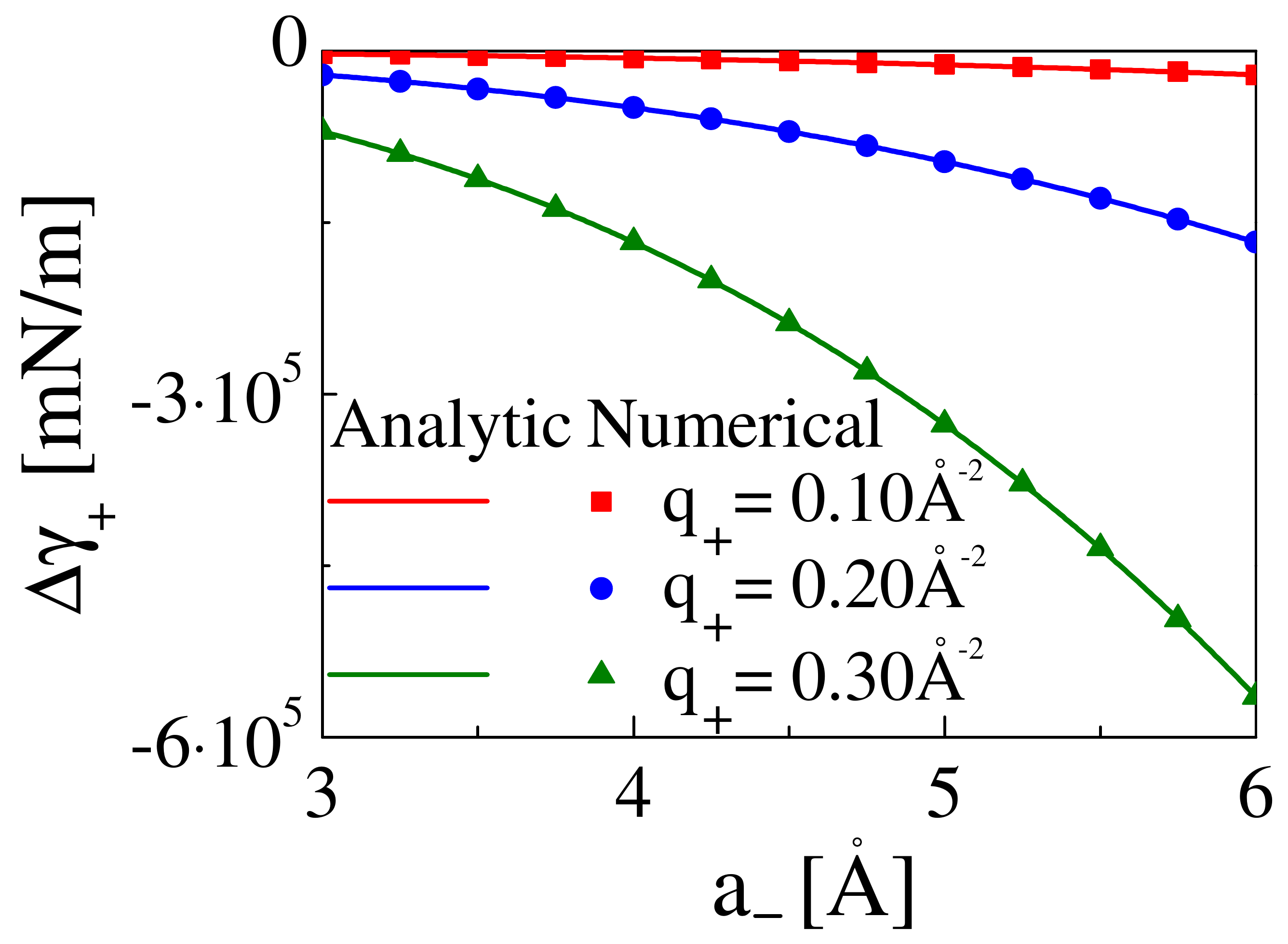}}
\caption{
\textsf{The excess surface tension $\Delta\gamma_{+}$ at the anode as a function of the ion size $a_-$. The bulk ion concentration is $c_b=1\rm mM$ and the cation size is $3\,{\rm \AA}$. The solid lines represent analytical calculations from Eq.~(\ref{e10}), while the data points are obtained via numerical calculations from Eqs. (\ref{e1}) - (\ref{e4}).
}}
\label{fig2}
\end{figure}

Figure \ref{fig2} shows the excess in surface tension $\Delta\gamma_{+}=\gamma_+-\gamma_0$ on the gel/anode interface as a function of anion size $a_{-}$ after numerically solving Eqs.~(\ref{e1})-(\ref{e4}). We can see that when an electric field is applied, the accumulation of anions introduces a negative contribution to the surface tension (instability) at the gel/anode interface. Hence, the gel tends to expand at the anode side, which is known as an electro-wetting effect. Moreover, the absolute value of the surface tension $\Delta\gamma_{+}$ is always an increasing function both of the ion size and charge density.

The derived theoretical framework allows us to deduce an analytical expression of $\Delta\gamma_{+}$ in the large $q$ limit. Under a large electric field, an electric double-layer is formed near the anode (and similarly near the cathode), and most of the voltage drop occurs in this double layer, which is shown in Fig.~1 in the supplemental material. We then assume that in the localized electric double-layer at the anode, the anions are closely packed, $c_{-}=1/a_{-}^3$, while cations are completely excluded, $c_{+}=0$. Considering charge neutrality, we can estimate the electric double-layer thickness to be $L_e^+=q_+ a_-^3$. Then, a simplified explicit expression of the electrostatic potential, $\Psi(z)$ is obtained,

\be
\label{e7}
\begin{aligned}
\Psi(z)=&\frac{2\pi l_{\rm B}}{a_{-}^3}\Big(z+\frac{1}{2}L\Big)^2-4\pi l_{\rm B} q_{+} \Big(z+\frac{1}{2}L\Big)\\
&+2\pi l_{\rm B} q_{+}^2a_{-}^3
\end{aligned}
\ee
for $0<|z+L/2|<L_e^+$
and $\Psi=0$ for $-L/2+L_e^+<z<0$.
In addition, the free-energy $F_{\rm elec}$ can be calculated from Eq.~(\ref{e1}).
\be
\label{e8}
\begin{aligned}
\frac{F_{\rm elec}}{k_{\rm B}T}=-\frac{4\pi}{3}l_{\rm B} (a_{-}q_{+})^3-q_{+}\tilde\mu_{-} ,
\end{aligned}
\ee
and the surface free-energy becomes

\be
\label{e9}
\begin{aligned}
\frac{g(q_{+})}{k_{\rm B}T}=\frac{2\pi}{3}l_{\rm B} (a_{-}q_{+})^3.
\end{aligned}
\ee
Finally, the excess surface tension on the anode side is

\be
\label{e10}
\begin{aligned}
\frac{\Delta\gamma_{+}}{k_{\rm B}T}=-\frac{4\pi}{3}l_{\rm B} (a_{-} q_{+})^3
\end{aligned}
\ee
The above $\Delta\gamma_{+}$ expression is shown as solid lines in Fig. \ref{fig2} and is in excellent agreement with the numerical results (data points). A similar behavior close to the gel/cathode interface, $\Delta\gamma_{-}\sim (a_{+} q_{-})^3$, is obtained, by simply exchanging negative with the positive charges.

\begin{figure}[h]
{\includegraphics[width=0.6\textwidth,draft=false]{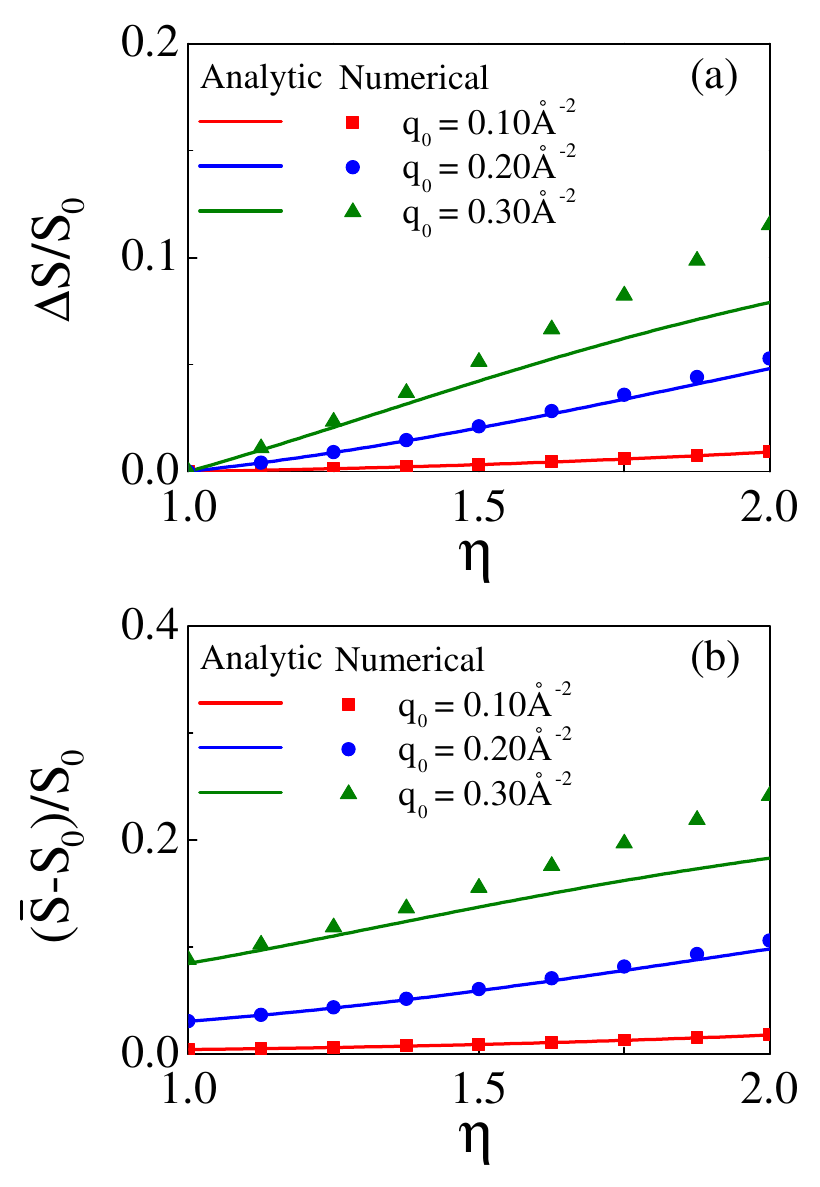}}
\caption{
\textsf{The asymmetric deformation of the PVC gel, $\Delta S/S_0$ and $(\overline{S}-S_0)/S_0$, as a function of the size ratio between anion and cation $\eta$. All solid lines are analytical results of Eqs.~(\ref{e12}) and (\ref{e13}), while the dotted data points are taken from the numerical calculations. (a) The dependence of $\Delta S/S_0$ on $\eta$ for three $q_0=Q/S_0$ values; (b) The effect of $\eta$ on $(\overline{S}-S_0)/S_0$ for three $q_0$ values. The other parameters are the bulk ions concentration $c_b=1\rm mM$, the elastic constant $k_b=k_s=15\,{\rm\AA}^{-2}$, $a_+=3\,{\rm \AA}$, $S_0=100\,{\rm \AA}^2$, $\varepsilon_r=10$, and $\l_{\rm B}=e^2/(4\pi\varepsilon_0\varepsilon_rk_BT)=55\,{\rm \AA}$, at room temperature, $T=300K$.
}}
\label{fig3}
\end{figure}

The electro-wetting induces an asymmetric deformation that can be further quantified by two parameters, the difference $\Delta S=S_{+}-S_{-}$ and the mean $\overline{S}=(S_{+}+S_{-})/2$. As shown in Fig.~\ref{fig3}a, for anions and cations, of equal size, $\eta=1$, the gel asymmetric deformation is zero, $\Delta S=0$, and independent of $q_0$, where $q_0=Q/S_0$ is the initial surface charge density. In addition, Fig.~\ref{fig3}a shows that the deformation $\Delta S$ increases as $\eta$ increases, indicating that the size ratio, $\eta$, plays a crucial role in controlling the gel asymmetric deformation. It is worth noticing that the $\eta$-induced gel deformation can be rather large, of order of $10\% - 20\%$. Moreover, the gel deformation can be further enhanced for $\eta \neq 1$ by increasing $q_0$. Figure~\ref{fig3}b indicates that the mean contact area $\overline{S}$ is an increasing function of $\eta$.

For large $q$, an analytical expression of $\Delta S^*$ and $\overline{S}^*$ can be obtained. Inserting the surface free-energy $g$, Eq.~(\ref{e9}), into the full free energy, $G$ of Eq.~(\ref{e5}) is written as
\be
\label{e11}
\begin{aligned}
\frac{G}{k_{\rm B}T}=&\frac{2\pi l_{\rm B} Q^3}{3}\left(\frac{a_{-}^3}{S_+^2}+\frac{a_{+}^3}{S_-^2}\right) \\
&+\frac{k_b}{2}(S_{+}-S_0)^2+\frac{k_b}{2}(S_{-}-S_0)^2+\frac{k_s}{2}(S_{+}-S_{-})^2
\end{aligned}
\ee
After rewriting $G$ as function of $\delta S_\pm$, where  $\delta S_\pm=S_\pm-S_0$, and expanding $G$ around $\delta S_\pm =0$ to the second order, the gel deformation $\Delta S^*$ and $\overline{S}^*$ can be obtained by minimizing $G$ with respect to $\delta S_\pm$,

\be
\label{e12}
\begin{aligned}
&\frac{\Delta S^*}{S_0}=\frac{1}{3}\frac{(\eta^3-1)}{\eta^3(\xi q_0^3+\kappa+1)+(1+2\kappa)/(\xi q_0^3)+\kappa+1}
\end{aligned}
\ee

\be
\label{e13}
\begin{aligned}
&\frac{\overline{S}^*-S_0}{S_0}=\frac{1}{6}\frac{\eta^3(2\xi q_0^3+2\kappa+1)+2\kappa+1}{\eta^3(\xi q_0^3+\kappa+1)+(1+2\kappa)/(\xi q_0^3)+\kappa+1}
\end{aligned}
\ee
where $\xi=4 \pi a_+^3 l_{\rm B}/(S_0 k_b)$ and $\kappa=k_s/k_b$.

The analytic expressions obtained from Eq.~(\ref{e12}) and (\ref{e13}) are shown in Fig. \ref{fig3} as solid lines, and agree well with the numerical results (dotted points). Note that the scaling law $\Delta S\sim\eta^3$ emphasizes the significant role played by the ionic size asymmetry on the asymmetric gel deformation. Such simple dependence of $\Delta S^*$ on $\eta$, $q_0$, and other experimental parameters found in Eq.~(\ref{e12}), can be verified experimentally.

Our analysis of $\Delta\gamma$ and $\Delta S$ indicates that the intriguing gel deformation is caused by accumulation of cations/anions having non-equal sizes. The applied electric field introduces a negative contribution to the surface tension $\Delta\gamma$.  In our model, $\Delta\gamma$ can compete with volume contributions like elasticity of the gel film because $\Delta\gamma$ is huge, of the order of $10^5$ mN/m and the PVC gel is soft due to the added plasticizers. Then, the absolute value of  $\Delta\gamma$ increases with ion size, while the gel volume remains constant during the deformation. Hence, the gel tends to spread at the anode where the larger size anions accumulate. Finally, the finite amount of asymmetric deformation is determined by the competition between the surface tension and gel elastic energy.

In addition, a very large apparent dielectric constant was observed in the PVC gel's experiments~\cite{Ali2011}. We note that such large values have been observed in the past for electrolytes~\cite{Colby2013},and serve as evidence for the existence of free ions inside the material. A more detailed consideration of the large values of the dielectric constant is presented in the supplemental material, where a simple analytic model relates these large dielectric values to the existence of free ions within the PVC gel. More precisely, its large value is determined by the ratio between the macroscopic film thickness (on the order of millimeters) and microscopic electric double-layer thickness (on the order of nanometers).

In summary, we developed a theoretical model unveiling the fundamental mechanism of the electro-activity of PVC gels. The combination of the electro-wetting effect and the ionic saturation is considered in this model. Furthermore, explicit analytic results related to the material deformation as seen in the experiments are presented. The theory shows that the size ratio between the mobile anions and cations plays a crucial role in determining the electro-induced deformation. An important conclusion is that increasing this ionic size ratio enhances the gel electro-actuation, which has yet to be verified in experiments. Finally, we hope that the physical concepts as explored in our model will provide insight into the design and applications of polymer actuation materials.

\textit{\textbf{Acknowledgements}}: This work was supported in part by Grant No. 21822302 of the National Natural Science Foundation of
China (NSFC), and by the NSFC-ISF
Research Program, jointly funded by the NSFC under Grant
No. 21961142020 and the Israel Science Foundation (ISF)
under Grant No. 3396/19.
We also acknowledge support from the Fundamental Research Funds for the Central Universities (China),
and the Israel Science Foundation (ISF) under Grant No. 213/19.

\bigskip\bigskip

\clearpage


\clearpage

\newpage
\vskip 0.5truecm
\centerline{for Table of Contents use only}
\centerline{\bf Enhanced electro-actuation in dielectric elastomers:}
\centerline{\bf the non-linear effect of free ions}
\centerline{\it Bin Zheng, Xingkun Man, David Andelman, and Masao Doi}

\begin{figure}[h]
{\includegraphics[width=0.6\textwidth,draft=false]{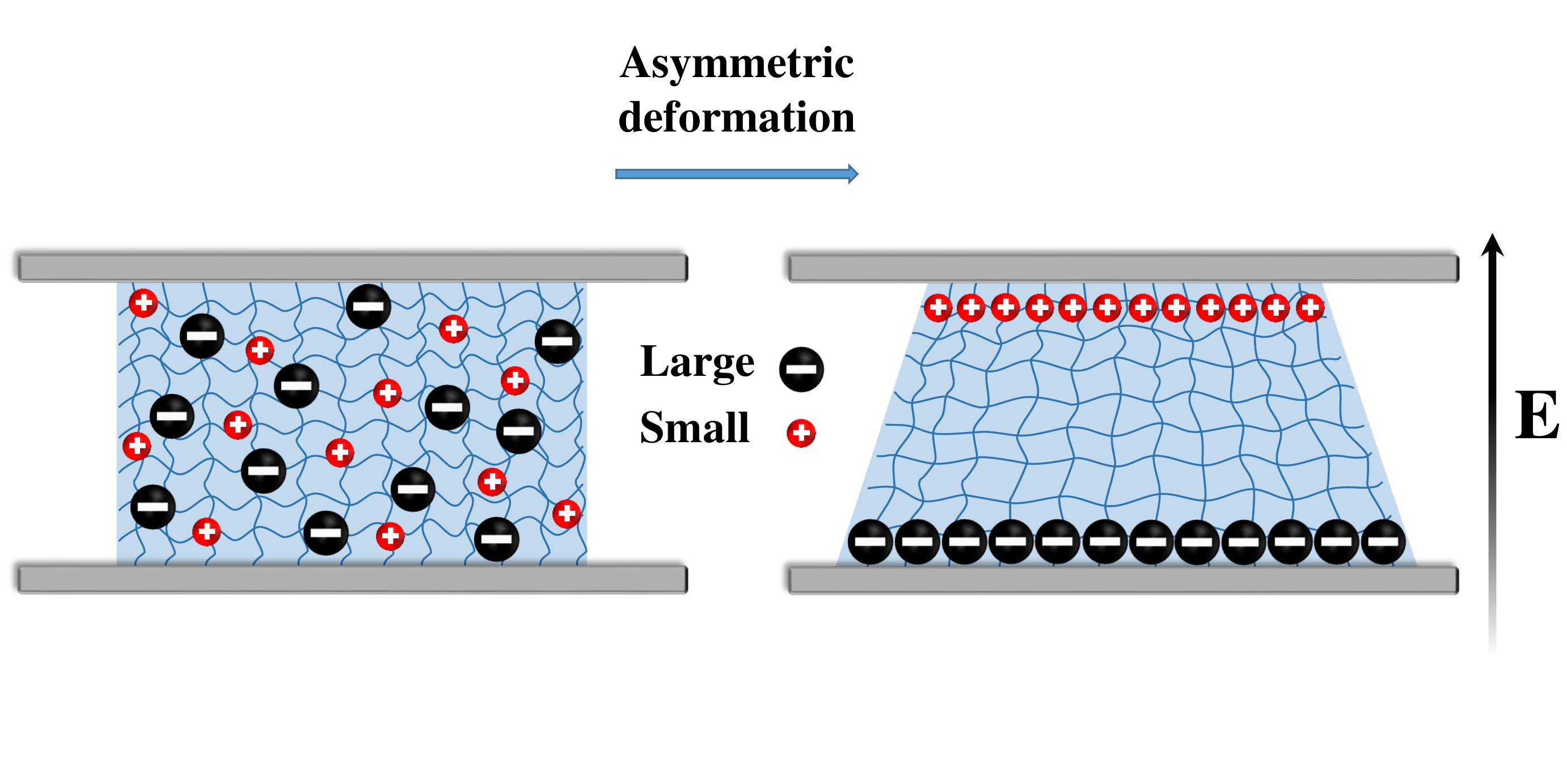}}
\end{figure}

\end{document}



\title{Supplemental Material for \\ Enhanced electro-actuation in dielectric elastomers: \\ the non-linear effect of free ions}
\author{Bin Zheng}
\affiliation{Raymond and Beverly Sackler School of Physics and Astronomy,
Tel Aviv University, Ramat Aviv 69978, Tel Aviv, Israel}
\author{Xingkun Man}
\email{manxk@buaa.edu.cn}
\affiliation{Center of Soft Matter Physics and its Applications, and School of
Physics and Nuclear Energy Engineering, Beihang University, Beijing 100191, China}
\author{David Andelman}
\email{andelman@post.tau.ac.il}
\affiliation{Raymond and Beverly Sackler School of Physics and Astronomy,
Tel Aviv University, Ramat Aviv 69978, Tel Aviv, Israel}
\author{Masao Doi}
\email{masao.doi@buaa.edu.cn}
\affiliation{Center of Soft Matter Physics and its Applications, and School of
Physics and Nuclear Energy Engineering, Beihang University, Beijing 100191, China}

\maketitle

\noindent\textit{\textbf{1. The Free Energy}}

The free energy is written as,
\be
\label{e1}
\begin{aligned}
\frac{F_{{\rm elec}}}{k_{\rm B}T}=&\int{\rm d^3}r \left\{-\frac{1}{8\pi l_{\rm B}}(\nabla\Psi )^2+(c_{+}-c_{-})\Psi
\right. \\
&\left. -\tilde\mu_{+}c_{+}-\tilde\mu_{-}c_{-}+c_{+}{\rm ln}(c_{+}a_{+}^3)+c_{-}{\rm ln}(c_{-}a_{-}^3)  \right. \\
&\left.+a_{{\rm gel}}^{-3}(1-c_{+}a_{+}^3-c_{-}a_{-}^3){\rm ln}(1-c_{+}a_{+}^3-c_{-}a_{-}^3)\right\}.
\end{aligned}
\ee
The first two terms account for the electrostatic energy, where $l_{\rm B}=e^2/(4\pi\varepsilon_0\varepsilon_rk_BT)$ is the Bjerrum length, $\varepsilon_r$ is the dielectric constant of the gel in the absence of ions, and  $\Psi={e\psi}/{k_{\rm B}T}$ is the dimensionless electrostatic potential. The third and fourth terms account for the chemical potential of the free ions, where $\tilde\mu_\pm={\mu_\pm}/{k_{\rm B}T}$ is the dimensionless chemical potential and $c_\pm(r)$ is the cation and anion number density. Because the thickness of the gel film, $L$, is large, the film is considered to be coupled to a bulk reservoir (inner film) of free ions via the ion chemical potential, $\tilde\mu_\pm$. The fifth and sixth terms are the entropy of the cations and anions. Finally, the last term in Eq.~(\ref{e1}) is the entropy of the gel molecules acting as a solvent. This solvent entropy is responsible for the saturation effect of the free ions that  accumulate near the electrodes\cite{david1997,david2000}.

Under large electric fields, due to the anion saturation at the gel/anode interface, the amount of cations and solvent molecules at the anode is negligible. Similarly, the amount of the anions and solvent molecules is negligible at the gel/cathode interface. For convenience purposes only, we can then decouple the last term in Eq.~(\ref{e1}) and write it as the sum of two separated steric terms. In addition, the gel size $a_{\rm gel}$ is taken the same as the size of the dominant ion respectively. The first term contributes only at the cathode, while the second one contributes only at the anode.
\be
\label{e2}
\begin{aligned}
\frac{F_{{\rm elec}}}{k_{\rm B}T}=&\int{\rm d^3}r \left\{-\frac{1}{8\pi l_{\rm B}}(\nabla\Psi )^2+(c_{+}-c_{-})\Psi
\right. \\
&\left. -\tilde\mu_{+}c_{+}-\tilde\mu_{-}c_{-}+c_{+}{\rm ln}(c_{+}a_{+}^3)+c_{-}{\rm ln}(c_{-}a_{-}^3)  \right. \\
&+a_{+}^{-3}(1-c_{+}a_{+}^3){\rm ln}(1-c_{+}a_{+}^3)+a_{-}^{-3}(1-c_{-}a_{-}^3){\rm ln}(1-c_{-}a_{-}^3)\bigg\}.
\end{aligned}
\ee

\bigskip
\noindent\textit{\textbf{2. The Ion Distribution and Electrostatic Potential inside the PVC Gel}}

\begin{figure}[h]
\centering
{\includegraphics[width=0.9\textwidth,draft=false]{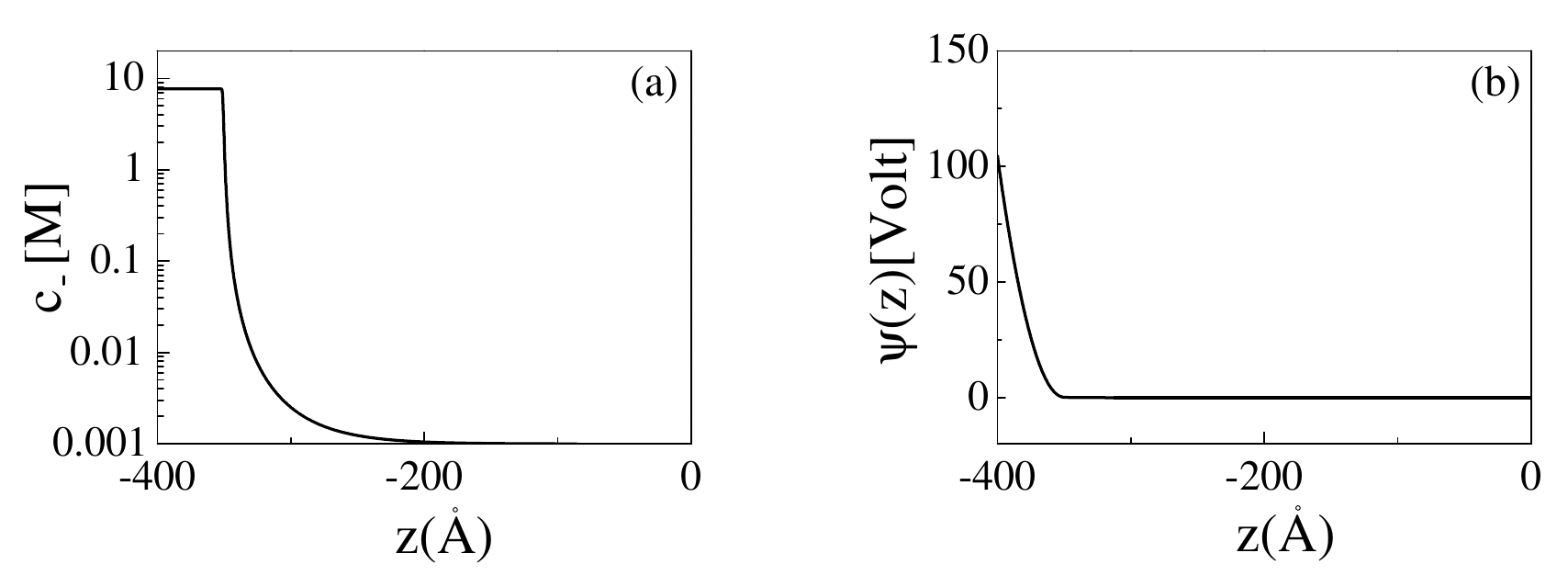}}
\caption{
\textsf{Electrostatic profiles for half of the system (anode side) as a function of $z$, the distance measured from the mid-plane between the electrodes. The anode is located at $z=-L/2=-400\,{\rm \AA}$. The anion distribution and electrostatic potential are plotted, respectively, in (a) and (b). The other parameters are $\eta=2.0$, $\Delta S/S_0= 0.115$, $(\overline{S}-S_0)/S_0=0.241$, $q_0=Q/S_0=0.3\,{\rm \AA}^{-2}$, $a_{+}=3\,{\rm \AA}$, and $c_b=1\rm mM$. Note that for a better presentation, we use semi-log plot in (a).
}}
\label{figs1}
\end{figure}

The numerical results presented in Fig.~\ref{figs1} show that anions are closely packed on the anode side with a number density $c_-=1/a_{-}^3$. As a result, there are almost no cations in that region. Correspondingly, there are closely packed cations and no anions on the cathode side. This is due to the high voltage (electric field on the order of 1kV/mm) applied across the electrodes. We assume charge neutrality of the entire system, {\it i.e.}, the total amount of the anions (cations) in the gel is equal to the charge number on the anode (cathode). As a result, we can estimate the thickness of the close-packed charge layer to be $L_e^+=q_{+}a_{-}^3$.

\bigskip
\noindent\textit{\textbf{3. Dielectric Constant}}

The other interesting behavior of the PVC gel is its large apparent dielectric constant, $\varepsilon_{{\rm app}}$, which is about $10^3-10^4$ times larger than the dielectric constant of the PVC and the embedded plasticizer~\cite{Ali2011}. We show how the  free ions play a crucial role in explaining this intriguing behavior.

To estimate the apparent dielectric constant, we model the system as a two-plate capacitor of thickness $L$. Around each electrode there is an electric double-layer of thickness $L_e^\pm$ coupled with the inner film of thickness $L-(L_e{^+}+L_e^-)$. We assume that the surface charge-densities on the electrodes is $\pm q$.
Thus, we can use the following relation for the capacitance,
\be
\label{e3}
\begin{aligned}
C=eq\frac{S}{V}=\varepsilon_0\varepsilon_{{\rm app}}\frac{S}{L}  ,
\end{aligned}
\ee
where the applied voltage $V=V_e+V_b$ has two contributions: a bulk part, $V_b$, and an electric double-layer part, $V_e$.
And $V$ can be approximately written as

\be
\label{e4}
\begin{aligned}
V=V_e+E_b L_b\approx\frac{1}{2}E_0(L_e^++L_e^-)+E_b L   ,
\end{aligned}
\ee
where $L_e^\pm$ was shown to be the electric double-layer thickness and $L_b=L-(L_e^{+}+L_e^{-})$ is the inner film thickness.  The electric field in the bulk is denoted as $E_b$, while $E_0\equiv eq/(\varepsilon_0\varepsilon_r)$ is the bulk electric field in the absence of free ions inside the gel. From Eqs.~(\ref{e3}) and (\ref{e4}), we obtain an explicit expression of the apparent dielectric constant, $\varepsilon_{{\rm app}}$,
\be
\label{e5}
\begin{aligned}
\frac{1}{\varepsilon_{{\rm app}}}\simeq\frac{1}{\varepsilon_r}\left(\frac{L_e^++L_e^-}{2L}+\frac{E_b}{E_0}\right).
\end{aligned}
\ee

\begin{figure}[h]
{\includegraphics[width=0.4\textwidth,draft=false]{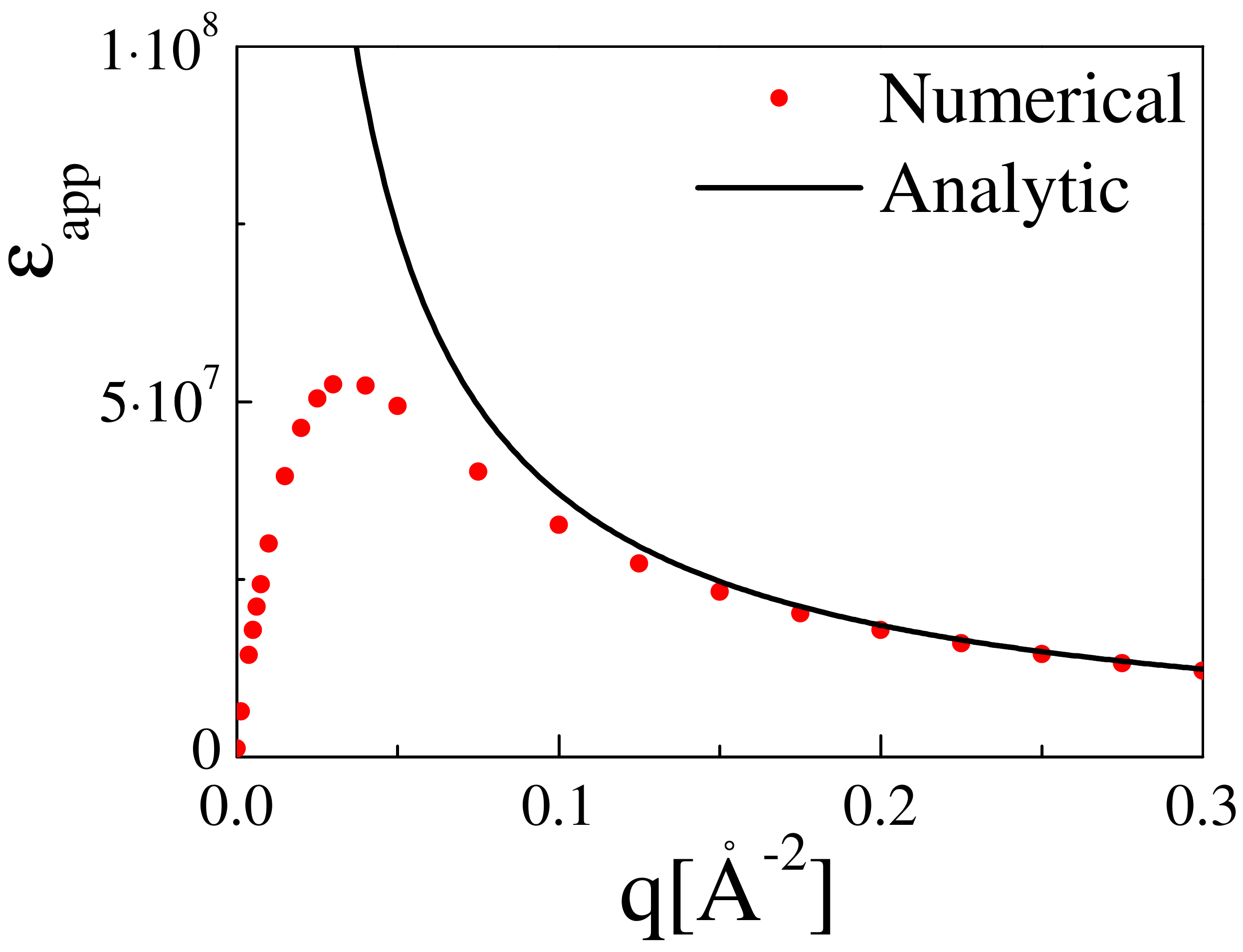}}
\caption{
\textsf{The apparent dielectric constant $\varepsilon_{{\rm app}}$ as a function of the surface charge density, $\boldmath{q=Q/S}$ on a semi-log plot. The other parameters are $c_b=1\rm mM$, $a_\pm=3\,{\rm \AA}$, $L=1{\rm mm}$ and $\varepsilon_r=10$.
}}
\label{figs2}
\end{figure}

Equation~(\ref{e5}) shows that indeed such free ions are at the origin of this special dielectric behavior. When there are no free ions, $\varepsilon_{{\rm app}}=\varepsilon_{{\rm r}}$, where $\varepsilon_{{\rm r}}$ includes the plasticizer and PVC contributions. The large $\varepsilon_{{\rm app}}$ values occur only when free ions are added, leading to a significant reduction of the bulk electric field, $E_b$. Moreover, the increase of the film thickness $L$ leads to a large apparent dielectric constant, $\varepsilon_{{\rm app}}$.

We further estimate the value of $\varepsilon_{{\rm app}}$ by ignoring the bulk electric field ($E_b=0$). For small $q$ near $0$, the electric double-layer thickness $L_e^\pm$ is of the order of the Debye length $\lambda_{\rm D}$. Then $\varepsilon_{{\rm app}}\simeq\varepsilon_{{\rm r}} L/\lambda_{\rm D}$ is independent of $q$ and is much larger than $\varepsilon_{{\rm r}}$. For large $q$, $L_e^\pm=q a_\pm^3$ and $V_e$ is obtained from the explicit expression of $\Psi(z)$, Eq.~(7) in the paper.
\be
\label{e6}
\begin{aligned}
\frac{eV_e}{k_{\rm B}T}=2\pi l_{\rm B} q^2(a_{-}^3+a_{+}^3) ,
\end{aligned}
\ee
yielding an approximate analytical expression of
\be
\label{e7}
\begin{aligned}
\varepsilon_{\rm{app}}=2\varepsilon_r L (a_-^3+a_+^3)^{-1} q^{-1}.
\end{aligned}
\ee

The dependence of $\varepsilon_{{\rm app}}$ on the surface charge-density $q$ is shown in Fig. \ref{figs2}. When the surface charge-density $q$ increases gradually from zero, the apparent dielectric constant $\varepsilon_{{\rm app}}$ first increases, reaching a maximum that is followed by a decrease.
